\documentclass[]{acmart}
\usepackage{multirow}
\usepackage{booktabs}
\AtBeginDocument{%
  }

\begin{document}
\title{Modifying AI, Enhancing Essays: How Active Engagement with Generative AI Boosts Writing Quality}
\renewcommand{\shorttitle}{How Active Engagement with Generative AI Boosts Writing Quality}

\author{Kaixun Yang}
\affiliation{%
  \institution{Monash University}
  \city{Melbourne}
  \country{Australia}}
\email{Kaixun.Yang1@monash.edu}

\author{Mladen Rakovi\'{c}}
\affiliation{%
  \institution{Monash University}
  \city{Melbourne}
  \country{Australia}}
\email{Mladen.Rakovic@monash.edu}

\author{Zhiping Liang}
\affiliation{%
  \institution{Monash University}
  \city{Melbourne}
  \country{Australia}}
\email{Zhiping.Liang@monash.edu}

\author{Lixiang Yan}
\affiliation{%
  \institution{Monash University}
  \city{Melbourne}
  \country{Australia}}
\email{Lixiang.Yan@monash.edu}

\author{Zijie Zeng}
\affiliation{%
  \institution{Monash University}
  \city{Melbourne}
  \country{Australia}}
\email{Zijie.Zeng@monash.edu}

\author{Yizhou Fan}
\affiliation{%
  \institution{Peking University}
  \city{Beijing}
  \country{China}}
\email{fyz@pku.edu.cn}

\author{Dragan Ga\v{s}evi\'{c}}
\affiliation{%
  \institution{Monash University}
  \city{Melbourne}
  \country{Australia}}
\email{Dragan.Gasevic@monash.edu}

\author{Guanliang Chen}
\authornote{Corresponding author}
\affiliation{%
  \institution{Monash University}
  \city{Melbourne}
  \country{Australia}}
\email{Guanliang.Chen@monash.edu}

\renewcommand{\shortauthors}{Yang and Rakovi\'{c}, et al.}

\begin{abstract}
Students are increasingly relying on Generative AI (GAI) to support their writing—a key pedagogical practice in education. In GAI-assisted writing, students can delegate core cognitive tasks (e.g., generating ideas and turning them into sentences) to GAI while still producing high-quality essays. This creates new challenges for teachers in assessing and supporting student learning, as they often lack insight into whether students are engaging in meaningful cognitive processes during writing or how much of the essay's quality can be attributed to those processes. This study aimed to help teachers better assess and support student learning in GAI-assisted writing by examining how different writing behaviors, especially those indicative of meaningful learning versus those that are not, impact essay quality. Using a dataset of 1,445 GAI-assisted writing sessions, we applied the cutting-edge method, X-Learner, to quantify the \textit{causal} impact of three GAI-assisted writing behavioral patterns (i.e., seeking suggestions but not accepting them, seeking suggestions and accepting them as they are, and seeking suggestions and accepting them with modification) on four measures of essay quality (i.e., lexical sophistication, syntactic complexity, text cohesion, and linguistic bias). Our analysis showed that writers who frequently modified GAI-generated text—suggesting active engagement in higher-order cognitive processes—consistently improved the quality of their essays in terms of lexical sophistication, syntactic complexity, and text cohesion. In contrast, those who often accepted GAI-generated text without changes, primarily engaging in lower-order processes, saw a decrease in essay quality. Additionally, while human writers tend to introduce linguistic bias when writing independently, incorporating GAI-generated text—even without modification—can help mitigate this bias.
\end{abstract}



\maketitle
\section{Introduction} \label{sec:intro}
In education, writing is a prevalent pedagogical practice employed by educators to facilitate students' learning \cite{Lea1998Student}. It can benefit students in various ways, such as developing analytical thinking \cite{grabe2014theory}, identifying knowledge gaps \cite{flower1981cognitive}, and enhancing communication skills \cite{langer1987writing}. In a well-known cognitive process model of writing \cite{flower1981cognitive}, writing is understood as a complex task that requires the coordination of three core cognitive processes: \textit{Planning}, \textit{Translating}, and \textit{Reviewing}. During the \textit{Planning} phase, writers engage in generating ideas, setting goals, and organizing their thoughts. This stage involves the decision-making process regarding what message the writer wants to communicate, the goals they hope to achieve, and the structure they will use to present their ideas logically. The \textit{Translating} phase follows, where writers take the ideas formed in the planning stage and convert them into written language. This process involves selecting appropriate vocabulary, constructing grammatically correct sentences, and ensuring the text is coherent and flows smoothly. Finally, in the \textit{Reviewing} phase, writers critically assess their draft by reading through the text to evaluate its overall quality. This stage involves identifying potential problems, such as unclear wording, weak arguments, or grammatical errors, and making necessary revisions. Writers may adjust content, refine style, or improve accuracy to enhance the effectiveness of their writing. In education, students are expected to actively engage in all three stages to achieve meaningful learning via writing tasks \cite{flower1981cognitive}. 

With the advent and increasing popularity of Generative AI (GAI), more and more higher education institutes have embraced GAI to support their teaching and learning, and GAI-assisted writing is now prevalent among students in higher education \cite{jin2024generative}. In GAI-assisted writing, students can delegate some of the core rhetorical and cognitive workload to GAI and receive help in content creation, improving creativity, and enhancing efficiency \cite{knowles2022human}. However, the use of GAI in writing introduces challenges for educators in supporting and assessing students' learning, as tasks that are originally undertaken by students to achieve meaningful learning are now performed by GAI. Specifically, students are traditionally responsible for independently planning, translating, and revising their work. With GAI support, students can now use GAI to generate ideas, convert those ideas into written text, and even request GAI to directly revise their written works. Unlike traditional writing, where the final product reflects the student's individual effort, GAI-assisted writing produces a combination of student-created and AI-generated content. Consequently, traditional writing assessment methods, which often focus on evaluating the quality of students' written product, might not be suitable in GAI-assisted contexts. Therefore, we argue it is crucial to understand the relationship between students' different GAI-assisted writing behaviors and the quality of the written essay. This will help determine whether improvements in essay quality are primarily driven by the capabilities of GAI or by meaningful learning undertaken by students during the writing process. In addition, it will guide educators in assessing whether more advanced assessment methods (e.g., considering students' in-process writing logs \cite{leijten2013keystroke}) are even more demanded than ever in this new context. Several studies have been conducted to assess whether essay quality improves when written with GAI \cite{dhillon2024shaping, luther2024teaming}. However, these studies were often limited in that they only investigated the overall impact of using GAI to support writing by comparing two groups: one with GAI-powered writing assistance and the other without, shedding little light on how different GAI-assisted writing behaviors influence writing quality. Such a binary comparison (i.e., with or without GAI) fails to account for differences in how students interact with GAI, which can affect both the writing process and outcomes. Moreover, it does not clarify whether meaningful learning occurs in the GAI-assisted context, as high-quality essays may simply result from GAI doing much of the work.

In GAI-assisted writing, it is common for students to directly incorporate GAI-generated text into their writing \cite{lee2022coauthor}. However, previous research has demonstrated that Large Language Models (LLMs, a main subset of GAI) can carry and amplify significant biases (e.g., gender, racial, and socioeconomic) \cite{kotek2023gender}. These biases arise from the training datasets, which often contain historical and societal inequities, reflecting real-world disparities in the representation of different genders, racial groups, and cultures \cite{acerbi2023large}. This raises concerns within the educational community, as linguistic biases, such as the use of terms like "\textit{female nurse}" or "\textit{male doctor}," embedded in LLMs may lead to biased outputs, which could be incorporated into students' work. A recent study underscored the risk of LLMs perpetuating harmful stereotypes through automated writing tools, potentially influencing students' writing styles and perspectives unintentionally \cite{wambsganss2023unraveling}. Therefore, in addition to traditional measures of writing quality such as coherence, syntax, and vocabulary \cite{isaacson1988assessing}, it is equally important to take the linguistic bias into account to measure writing quality and investigate its relationship with different GAI-assisted writing behaviors.

To address these gaps, this study aimed to answer the following Research Question (RQ): \textbf{What writing behaviors contribute to the quality of written products in the setting of GAI-assisted writing?} In this study, we chose a publicly available dataset, CoAuthor \cite{lee2022coauthor}, containing 1,445 GAI-assisted writing sessions, including both final outputs and log trace events from the GAI-assisted writing process. We focused on three types of GAI-assisted writing behavioral patterns that may indicate meaningful learning or its absence, namely \textit{only seeking suggestions from GAI but not accepting them}, \textit{seeking suggestions from GAI and accepting them as they are}, and \textit{seeking suggestions from GAI and accepting them with modification}. To assess the quality of the written products, we selected three measures frequently used in previous writing research, namely \textit{lexical sophistication}, \textit{syntactic complexity}, and \textit{text cohesion} \cite{crossley2020linguistic}. Additionally, we incorporated \textit{gender bias} (e.g., "\textit{She was too emotional to make a rational decision.}") as a representative of linguistic biases to further evaluate the quality of the written products. We aimed to identify causal relationships between GAI-assisted writing behaviors and the resulting quality in the written products to address the RQ so that such insights can be directly employed to guide real-world pedagogical writing practices. While randomized controlled trials (RCTs) are often considered the gold standard for establishing causal relationships \cite{cartwright2010randomised}, practical constraints such as ethical concerns and implementation challenges \cite{hariton2018randomised} make RCTs not always feasible. In our study, the variety of writing behaviors (e.g., accepting or modifying GAI suggestions) and contexts (e.g., different writing topics) makes designing RCTs particularly difficult. Instead, we applied causal modeling, a statistical method designed to uncover and understand cause-and-effect relationships using observational data \cite{feder2022causal}. Specifically, we treated the GAI-assisted writing behavioral patterns displayed by writers as \texttt{treatments} and the four adopted essay quality measures as \texttt{outcomes}. Then, we encoded all treatments and outcomes into a Directed Acyclic Graph to model the causal relationships among them while conditioning on factors such as writers' first-language backgrounds and the genre of a writing task. Then, the causal relationships were identified using the widely-used back-door criterion \cite{maathuis2015generalized} and estimated through the state-of-the-art X-learner algorithms \cite{kunzel2019metalearners}. Further details can be found in Section \ref{sec:method}. Through extensive analysis, this study contributed with the following major findings: (i) The direct use of GAI-generated text, which suggests minimal learning by the writer, often reduces the quality of an essay in terms of lexical sophistication, syntactic complexity, and cohesion. However, actively revising GAI-generated text, which involves meaningful learning during the writing process, can significantly improve essay quality across all three measures. (ii) When writing independently, human writers are likely to introduce linguistic bias into their text. By incorporating GAI-generated text—whether through revision or direct use—writers can produce essays of higher quality with reduced linguistic bias. (iii) Non-native English writers often benefit from GAI in improving the lexical sophistication and syntactic complexity of their essays. However, they tend to exhibit a higher degree of linguistic bias when writing primarily on their own or revising GAI-suggested text. (iv) Actively revising GAI-suggested text tends to improve the cohesion of essays more in creative writing, but has a lesser impact in argumentative writing.

\section{Related Work}
\subsection{Writing Assessment in Education} \label{sec:cm}
Writing plays a crucial role in education by promoting deeper cognitive engagement, encouraging critical analysis, and fostering the synthesis of ideas, all of which support the development of higher-order cognitive skills \cite{applebee1984writing}. \citet{emig1977writing} described writing as a unique mode of learning that enhances students' abilities in analysis, synthesis, and reasoning. 

Writing assessment is essential for offering targeted feedback that helps students reach desired learning outcomes and improve their writing skills through writing activities \cite{condon2004assessing}. Most research on writing assessment has focused on evaluating the final written product, i.e., \textit{product-based} assessment. One common approach involves assessing the linguistic features of written work \cite{taguchi2013linguistic}. Linguistic features of writing generally fall into three broad categories \cite{mcnamara2010linguistic}: lexical, syntactic, and cohesion. Specifically, \citet{crossley2020linguistic} emphasized the strong relationships between essay quality and these three aspects: lexical sophistication, syntactic complexity, and text cohesion. Specifically, (i) \textbf{Lexical sophistication} refers to the use of advanced vocabulary in a text and often serves as a key indicator of text quality \cite{crossley2020linguistic}. 
\citet{kyle2016relationship} underscored the importance of lexical sophistication in writing proficiency, particularly in the context of second language acquisition. Similarly, \citet{laufer1995vocabulary} showed that the lexical frequency serves as a reliable and stable measure effectively distinguishing between writers' proficiency levels. (ii) \textbf{Syntactic complexity} pertains to both the sophistication and variety of syntactic structures used \cite{lu2011corpus}. \citet{yang2015different} identified global sentence and T-unit complexity (i.e., the shortest grammatically complete sentence) as consistent predictors of writing scores across various topics. Similarly, \citet{ha2022syntactic} also found a significant positive relationship between global features, such as Mean Length of Sentence and Mean Length of T-unit, and writing scores across different topics. (iii) \textbf{Text cohesion} refers to how well different parts of the text are connected through lexical, semantic, and argumentative links, making it a crucial component of effective writing \cite{halliday2014cohesion}. \citet{crossley2016say} demonstrated that increasing text content and enhancing cohesion result in significant improvements in evaluations of writing quality.  \citet{crossley2010cohesion} also suggested that coherence is a key factor in determining overall essay quality. While \textit{product-based} assessment has been widely applied in existing writing research, \citet{boud2007reframing} highlighted that such assessments often prioritize the final outcome over the learning process, potentially neglecting the development of critical writing skills. We argue that this issue may be even more pronounced in the GAI-assisted writing context, where students may rely heavily on GAI to generate text, resulting in minimal engagement with the actual learning process. 

An alternative approach is \textit{process-based} assessment, which focuses on evaluating the methods and strategies students employ throughout the writing process, such as planning, translating, reviewing, and monitoring \cite{koppenhaver2010conceptual}. Existing studies often extract writing behaviors (e.g., between-word pauses, the number of writing bursts, and the frequency of revisions) from keystroke logs, using them as measurable indicators of writing strategies, cognitive processes, and essay quality. For example, \citet{conijn2019understanding} proposed mapping features from keystroke logs to higher-level cognitive processes like planning and revising. Similarly, \citet{baaijen2012keystroke} developed methods and measures for analyzing keystroke data with the aim of enhancing the correlation between keystroke patterns and cognitive processes. \citet{deane2015exploring} demonstrated that planning generally leads to more logically organized texts, while the revision phase is particularly valuable for improving grammar and coherence. However, these studies typically focus on human-only writing contexts and do not account for GAI-assisted writing. Rather than directly developing a \textit{process-based} assessment method for GAI-assisted writing, our goal is to first gain insights into how different writing behaviors impact essay quality, which can then inform the design of relevant process-based assessment methods tailored to GAI-assisted writing.

\subsection{Generative AI-assisted Writing}
Existing research on GAI-assisted writing is relatively limited. A few studies focused on developing GAI-assisted writing systems and gathering relevant datasets. For instance, \citet{coenen2021wordcraft} introduced Wordcraft, an AI-assisted editor designed for collaborative story writing using few-shot learning and conversational affordances. Another notable work by \citet{lee2022coauthor} presented CoAuthor, a dataset capturing interactions between 63 writers and four instances of GPT-3 across 1,445 writing sessions. Other research focused on analyzing writers' writing behaviors in GAI-assisted writing. For example, \citet{cheng2024evidence} proposed a methodology based on learning analytics to evaluate human writing processes in GAI-assisted writing, comparing writing behaviors across different groups by using the CoAuthor dataset (e.g., creative vs. argumentative writing). \citet{yang2024inkalgorithmexploringtemporal} applied time-series clustering to identify common behavior patterns in GAI-assisted writing, noting similarities with traditional peer collaborative writing. Further studies correlated human writing behaviors with writing performance. For example, \citet{shibani2023visual} introduced CoAuthorViz, a tool that visualizes keystroke logs from GAI-assisted writing and explored the correlations between various human writing behaviors and the quality of the final products. \citet{nguyen2024human} used Hidden Markov Models combined with hierarchical sequence clustering to analyze human-AI interactions in academic writing, finding that doctoral students engaging in iterative, highly interactive writing processes with AI tools tend to perform better. Additionally, since LLMs have been shown to exhibit a tendency to generate outputs that reflect societal stereotypes, prejudices, or imbalances based on factors such as gender, race, and culture \cite{navigli2023biases, kotek2023gender}, one study has started investigating whether these biases, particularly gender bias, transfer from LLMs into students' writing when using AI-assisted tools \cite{wambsganss2023unraveling}. 

While these studies identified behavioral patterns in GAI-assisted writing, they often fall short of establishing a clear connection between these patterns and the quality of the written output (e.g., \cite{yang2024inkalgorithmexploringtemporal, cheng2024evidence}) or only found correlations rather than causal relationships between writing behaviors and essay quality, thus being limited in guiding real-world pedagogical practices in GAI-assisted writing (e.g., \cite{shibani2023visual, nguyen2024human}). 
Our study distinguished itself from existing research by identifying causal relationships between human writing behaviors and different essay quality measures with the ultimate goal of offering more actionable insights for designing next-generation GAI-assisted writing systems that better support student learning. 

\subsection{Identifying Causal Relationships in Educational Research}
In educational research, randomized controlled trials (RCTs) are commonly used to establish causal relationships. For example, \citet{cantrell2016supplemental} examined the effectiveness of supplemental literacy instruction for elementary students struggling with reading through RCTs, revealing significant improvements in reading skills for students in the treatment group. \citet{hariton2018randomised} evaluated the impact of scholarships on high school graduation rates using RCTs, finding that scholarships substantially increased the likelihood of students graduating. However, RCTs are not always feasible due to practical limitations such as ethical concerns and challenges related to compliance and implementation \cite{hariton2018randomised}. Consequently, some researchers have turned to causal modeling to identify causal relationships, which employs statistical methods to infer causal relationships from observational datasets. For instance, \citet{Denton1981Causal} applied causal modeling to student-teacher data and found that performance objectives and learner diagnosis accounted for over half of the variance in instructional strategies. \citet{Parkerson1984Exploring} found that home environment and peer groups were significant predictors of educational achievement. Additionally, \citet{Wesson1988A} demonstrated that accurate measurement of student performance and structured instruction directly contributed to academic success.

To the best of our knowledge, no previous studies have attempted to explore the causal relationships between human writing behaviors and essay quality in the setting of GAI-assisted writing. Our research represents the first empirical investigation into these causal relationships within GAI-assisted writing tasks. By applying causal inference methods, we can provide valuable insights for educators, guiding them in adjusting assessment practices to consider GAI's impact on student writing, refining instructional strategies for effective AI tool use, designing AI systems suited to various writing contexts, and ultimately enhancing student learning through GAI-assisted writing.

\section{Methodology} \label{sec:method}

\subsection{Dataset}
The CoAuthor dataset\footnote{\url{https://coauthor.stanford.edu/}} is a publicly available dataset focused on GAI-assisted writing. It was developed with contributions from 63 writers recruited from a crowd-sourcing platform, each of whom participated in multiple sessions across 20 writing topics. Throughout the writing process, writers could request sentence suggestions from GPT-3, receiving up to five suggestions at any time. Writers had the flexibility to accept, reject, or first accept and then modify these suggestions to better fit their writing. The dataset consists of 1,445 writing sessions, including 830 creative writing sessions and 615 argumentative writing sessions. A detailed description of the dataset is provided in \cite{lee2022coauthor}. We acknowledge that this dataset was not crafted with real-world students. However, the data collection process involved clear writing instructions, and participants received appropriate training in the writing tasks (i.e., creative and argumentative writing) that highly resembled those used in authentic educational scenarios \cite{ten2024instruction}. Therefore, we believe that this dataset can be used to address our RQ and offer valuable insights for authentic educational scenarios. It is worth noting that several educational studies have already recognized the value of this dataset and incorporated it into their research \cite{cheng2024evidence, shibani2023visual, yang2024inkalgorithmexploringtemporal}. 

\subsection{Conceptual Causal Model}
In causal modeling \cite{petersen2014causal}, the terms \textbf{treatments}, \textbf{outcomes}, and \textbf{confounders} are fundamental concepts that help define the relationships being studied. The \textbf{treatment} refers to the variable or intervention that is being applied or manipulated to observe its effect on another variable. The \textbf{outcomes} are the variables that researchers are interested in measuring to see if they are affected by the treatment. The \textbf{confounders} are variables that influence both the treatment and the outcome, potentially distorting the apparent effect of the treatment on the outcome. The key goal of causal modeling is to establish the relationship between treatment and outcome while conditioning on confounders. 

We argued that GAI-assisted writing processes can be somewhat analogous to the cognitive model of writing described in Section \ref{sec:intro}. In the \textbf{Planning} phase, writers generate ideas, set goals, and organize their thoughts, deciding what to convey and how to structure their work. We argued that when writers seek suggestions for GAI, they are essentially gathering ideas from the GAI, which aligns with the \textbf{Planning} stage in traditional writing. In the \textbf{Translating} phase, writers transform their ideas into written language, selecting appropriate words, forming sentences, and ensuring clarity and coherence. We argued that when writers incorporate GAI-generated text into their work, they are using the GAI to convert ideas into written language, aligning with the \textbf{Translating} stage in the traditional writing process. In the \textbf{Reviewing} phase, writers evaluate their text, assess its quality, and make revisions by identifying areas for improvement in content, style, or accuracy. We argued that when writers modify AI-generated text to better match their tone or intent, they are engaging in the \textbf{Reviewing} stage. In a recent study on GAI-assisted writing, \citet{yang2024inkalgorithmexploringtemporal} identified distinct patterns of GAI-assisted writing behaviors (e.g., writers who mostly prefer independent writing without GAI assistance) using a clustering approach. Inspired by this study, we are interested in the following three behavioral patterns: (i) writers sought suggestions from GAI but seldom accept such GAI-generated text for their writing (either due to their distrust of GAI or any other reasons); (ii) writers who tended to blindly trust GAI, i.e., they often sought suggestions from GAI and accepted them as they were; and (iii) writers who partly trusted GAI, i.e., they often sought suggestions from GAI and accepted them with modification. Correspondingly, we defined three treatments in our study: (T1) seek suggestions -> not accept; (T2) seek suggestions -> accept without revision; and (T3) seek suggestions -> accept and then revise. It should be noted that these three behavioral patterns (or treatments) are likely indicative of different levels of cognitive engagement during the writing process. Specifically, compared to T1 and T2, T3 entailed the most active engagement from writers as writers needed to critically evaluate the relevance of the suggested text and then modify them as needed. Causal modeling methods \cite{alipourfard2018using} typically require the treatment to be a binary variable (i.e., receiving or not receiving the treatment). Therefore, we transformed the patterns of writing behaviors into binary variables by classifying them as either above or below the median value of occurrence frequency of a behavioral pattern in the writing process.

We measured the quality of essay products from four different perspectives, namely lexical sophistication, syntactic complexity, text cohesion, and linguistic bias, and used them as the outcome variables (i.e., Y1, Y2, Y3, and Y4) in the causal modeling process. We calculated each of these measures by borrowing well-studied metrics from relevant literature \cite{crossley2020linguistic, wambsganss2023unraveling}: (i) Lexical sophistication (Y1): Advanced Guiraud refers to an enhanced version of the Guiraud Index, a measure of lexical sophistication used in linguistics and language studies \cite{daller2003lexical}. A higher Advanced Guiraud value indicates greater lexical sophistication. (ii) Syntactic complexity (Y2): Mean Length of T-Unit (MLT) is a metric used in linguistics to analyze syntactic complexity and writing proficiency \cite{bardovi1992second}. A higher MLT value indicates greater syntactic complexity. (iii) Text cohesion (Y3): Semantic Overlap refers to the degree to which two or more words, phrases, sentences, or texts share similar meanings or convey related concepts. A higher semantic overlap indicates greater textual cohesion. (iv) Here, we studied linguistic bias from the perspective of gender bias, which is the most prevalent form of bias often found in LLMs \cite{wambsganss2023unraveling}. Specifically, the Genbit Score is a metric used to measure gender bias in language datasets \cite{sengupta2021genbit}. A higher score reflects a greater degree of bias present in the text. 

In GAI-assisted writing, there can be various confounders impacting how writers make use of GAI and their subsequent essay products. Guided by previous studies, we studied the following five confounders (C1, C2, C3, C4, and C5), all of which might be important in GAI-assisted writing: (i) \textbf{Writing genre} (C1) refers to a specific category or style of writing, defined by its purpose, structure, and target audience \cite{grimmer2018evolution}. For instance, creative writing emphasizes emotional and artistic expression \cite{forgeard2013psychology}, while argumentative writing centers on logical reasoning and persuasion \cite{wingate2012argument}. (ii) Writing composition is shaped by a range of skills, including discourse-level language abilities, working memory, and transcription skills. These factors form a hierarchical structure that influences how individuals write, with language proficiency being a key factor in determining writing outcomes \cite{kim2020structural}. Therefore, a writer's \textbf{language background} (C2) should also be considered a potential confounding factor. (iii) \textbf{Writing topic} (C3) can be a potential confounding factor, as students write better essays on topics they find personally meaningful or enjoyable, compared to those they find less engaging \cite{boscolo2006writing}. (iv) \textbf{GPT temperature} (C4) and \textbf{Frequency penalty} (C5) settings are two potential confounding factors as well. Temperature adjusts the level of randomness or creativity in the GAI’s responses, while the frequency penalty discourages GAI from repeating the same words or tokens within a single response. That is, a high temperature results in more diverse and creative outputs, while a high frequency penalty decreases repetition, promoting a broader vocabulary.  

The details about model variables can be found in Table \ref{table1}. In causal modeling, it is impossible to account for all factors in reality, making it difficult to claim that the estimated effect is entirely free of bias. However, the selection of variables is strongly guided by existing study discussed above, which is expected to minimize potential bias. Additionally, we rigorously test the reliability of the effect estimates through multiple refutation checks, as outlined in Section \ref{sec:refu}.

\begin{table*}[hbt!]
\begin{center}
\caption{Conceptual causal model variable code, description, and identifiers.}
\label{table1}
\resizebox{1\textwidth}{!}{
\begin{tabular}{@{}lll@{}}
\toprule
Code & Pattern Description & Identifiers \\ \midrule
T1 & seek suggestions -> not accept & Number of attempts asking for GAI suggestions without subsequent acceptance \\
T2 & seek suggestions -> accept without revision & Ratio between the accepted GAI suggestions without modification and the total number of accepted GAI suggestions \\
T3 & seek suggestions -> accept with revision & Ratio between teh accepted GAI suggestions with modification and the total number of accepted GAI suggestions \\ \midrule
Y1 & Lexical sophistication & Advanced Guiraud \\
Y2 & Syntactic complexity & Mean Length of T-Unit \\
Y3 & Text cohesion & Semantic Overlap \\
Y4 & Gender bias & Genbit Score \\ \midrule
C1 & Writing genre & 1 for creative writing, 0 for argumentative writing \\
C2 & Writing topic & 20 different topics \\
C3 & First-language background & 1 for native English writers, 0 for non-native English writers \\
C4 & GPT temperature & 0.2, 0.3, 0.75, and 0.9 \\
C5 & GPT frequency penalty & 0, 0.5, and 1 \\ \bottomrule
\end{tabular}
}
\end{center}
\end{table*}

\begin{figure*}[hbt!]
\includegraphics[width=0.5\textwidth]{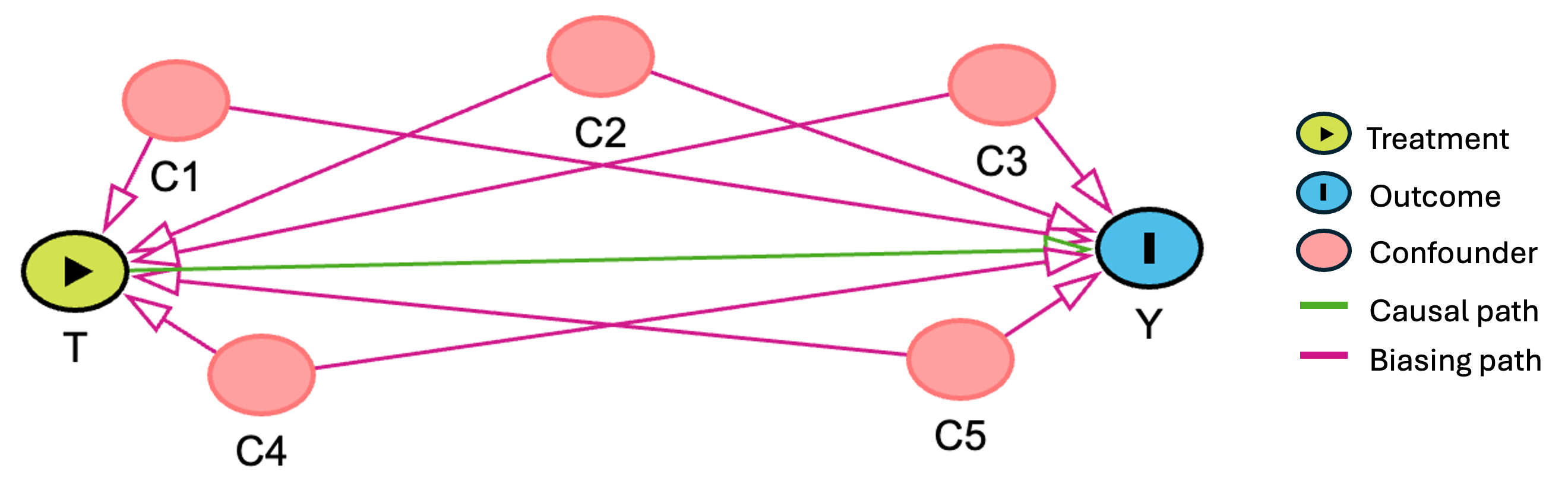}
\caption{Causal graph of GAI-assisted writing.} 
\label{fig1}
\end{figure*}

We then encoded the conceptual model as a Directed Acyclic Graph (DAG) \( G \equiv (V, E) \), where \( V \) represents a set of vertices containing all relevant variables, and \( E \) represents a set of directed edges connecting them. A directed edge \( A \rightarrow B \) indicates that \( A \) causes \( B \), meaning that altering the value of \( A \) while keeping everything else constant will affect the value of \( B \). The causal graph is shown in Figure \ref{fig1}. Our goal is to identify and measure the causal path while mitigating the influence of biasing paths  (i.e., pathways in DAG that introduce bias into the estimation of causal effects).

\subsection{Causal Effect Identification and Estimation}
Our objective was to evaluate the effect of a treatment on an outcome. However, in reality, we cannot observe counterfactual values, which are essential for calculating causal effects but are inherently unobservable. To address this, we rely on identification techniques and assumptions that reduce causal estimands into statistical estimands (i.e., estimated from observed data). We chose the widely-used back-door criterion \cite{maathuis2015generalized} for identification, which uses a graphical test to determine if adjusting for a set of variables (denoted as back-door adjustment set) is sufficient to identify causal estimands from observational data. Once we identify a suitable back-door adjustment set, the causal effects are measured by applying the Average Treatment Effect (ATE) and Individual Treatment Effect (ITE). The ATE represents the overall average causal effect of the treatment across the entire population. It quantifies the difference in outcomes between those who receive the treatment and those who do not. Mathematically, ATE is defined as:
\begin{equation}
\text{ATE} = E[Y(1)] - E[Y(0)]
\end{equation}

Where \( E[Y(1)] \) is the expected outcome if the entire population received the treatment, \( E[Y(0)] \) is the expected outcome if no one received the treatment.

ITE, on the other hand, examines the effect of a treatment on a specific individual, rather than an average effect across a population. Mathematically, ITE is expressed as:
\begin{equation}
\text{ITE}_i = Y_i(1) - Y_i(0)
\end{equation}

Where \( Y_i(1) \) represents the outcome for individual \(i\) if treated, \( Y_i(0) \) represents the outcome for individual \(i\) if not treated.

In our study, ATE and ITE estimations were conducted by the state-of-the-art X-learner algorithms \cite{kunzel2019metalearners}. The central concept of the X-learner is to leverage data from both treated and untreated groups to improve predictions about how each individual would respond to a treatment. Please refer to the digital appendix for descriptions about relevant methodological details \footnote{\url{https://github.com/CarsonYang518/LAK25-GAI-Writing-Causal}}.

\subsection{Evaluation} \label{sec:refu}
\noindent\textbf{Refutation Methods}. Since ground-truth estimates are not directly observed, we rely on refutation methods \cite{dowhy} to conduct robustness checks and sensitivity analyses of the estimates. These include: (i) Random Common Cause (RCC): introducing a random confounder into the model to assess how the estimated causal effect changes, testing its robustness; (ii) Placebo Treatment Refuter (Placebo): applying a placebo treatment to determine if the estimated causal effect is attributable to random chance; and (iii) Data Subset Refuter (DSR): re-estimating the causal effect on different subsets of the data to evaluate its stability. For all methods, a p-value greater than 0.05 indicates a successful test.

\noindent\textbf{SHapley Additive exPlanations (SHAP)}. To further investigate the extent to which different confounders impact ITE, we generated SHAP beeswarm plots \cite{antwarg2021explaining} to visualise the varying importance of confounders in the X-learner causal models. The beeswarm plot provides a visual summary of how the most important features in a dataset influence the model’s output. Each explanation is represented by a dot on a feature row, where the dot's x-position is based on the feature's SHAP value, indicating its impact on the prediction. Dots accumulate along the row to reflect density, and color is used to represent the original value of each feature. In our study, the model output is the ITE, and the features are confounders. More specifically, each dot represents an individual writing session within the dataset, with the colors indicating the value of a confounding variable associated with each data point.

\section{Results}
\begin{table*}[hbt!]
\begin{center}
\caption{Results of Average Treatment Effect estimation. For the Placebo, RCC and DSR refutations, the new ATE estimate is reported (denoted as Effect*), alongside the respective p-value and a p-value less than 0.05 indicates a failed test in causal modeling.}
\label{table2}
\resizebox{0.65\textwidth}{!}{
\begin{tabular}{@{}cc|c|cccccc@{}}
\toprule
\multirow{3}{*}{Treatment} & \multirow{3}{*}{Outcome} & \multirow{3}{*}{ATE} & \multicolumn{6}{c}{Refutations} \\ \cmidrule(l){4-9} 
 &  &  & \multicolumn{2}{c}{RCC} & \multicolumn{2}{c}{Placebo} & \multicolumn{2}{c}{DSR} \\ \cmidrule(l){4-9} 
 &  &  & Effect* & p-value & Effect* & p-value & Effect* & p-value \\ \midrule
T1 & Y1 & -0.065 & -0.065 & 0.96 & 0.002 & 0.86 & -0.062 & 0.84 \\
T1 & Y2 & -0.236 & -0.150 & 0.44 & 0.007 & 0.90 & -0.249 & 0.92 \\
T1 & Y3 & 0.001 & 0.007 & 0.22 & 0.000 & 0.86 & 0.002 & 0.84 \\
T1 & Y4 & -0.163 & -0.142 & 0.44 & -0.002 & 0.92 & -0.152 & 0.84 \\ \midrule
T2 & Y1 & -0.086 & -0.086 & 0.94 & -0.004 & 0.84 & -0.086 & 0.98 \\
T2 & Y2 & -0.869 & -0.862 & 0.92 & -0.024 & 0.86 & -0.889 & 0.80 \\
T2 & Y3 & -0.007 & -0.008 & 0.46 & 0.001 & 0.88 & -0.007 & 0.86 \\
T2 & Y4 & -0.012 & -0.007 & 0.54 & -0.001 & 0.86 & -0.013 & 0.90 \\ \midrule
T3 & Y1 & 0.102 & 0.102 & 0.86 & -0.002 & 0.94 & 0.101 & 0.90 \\
T3 & Y2 & 0.963 & 0.965 & 0.96 & -0.007 & 0.88 & 0.956 & 0.90 \\
T3 & Y3 & 0.008 & 0.008 & 0.68 & 0.000 & 1.00 & 0.008 & 0.94 \\
T3 & Y4 & -0.013 & -0.017 & 0.56 & 0.006 & 0.94 & -0.014 & 0.92 \\ \bottomrule
\end{tabular}
}
\end{center}
\end{table*}

\begin{figure*}[hbt!]
\includegraphics[width=0.65\textwidth]{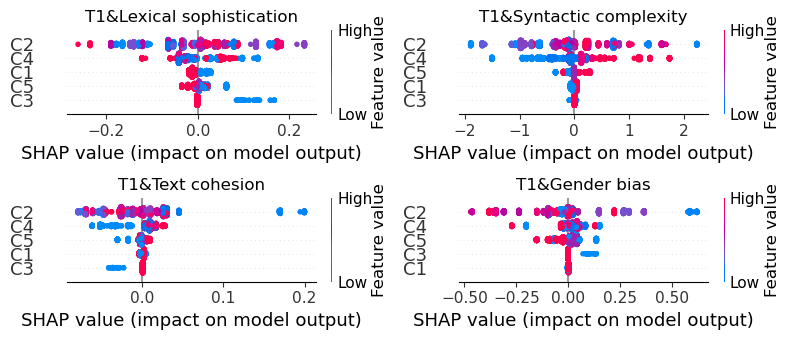}
\caption{Beeswarm plot of \texttt{seek suggestions -> not accept} (T1) \& outcomes.} 
\label{fig2}
\end{figure*}

\begin{figure*}[hbt!]
\includegraphics[width=0.65\textwidth]{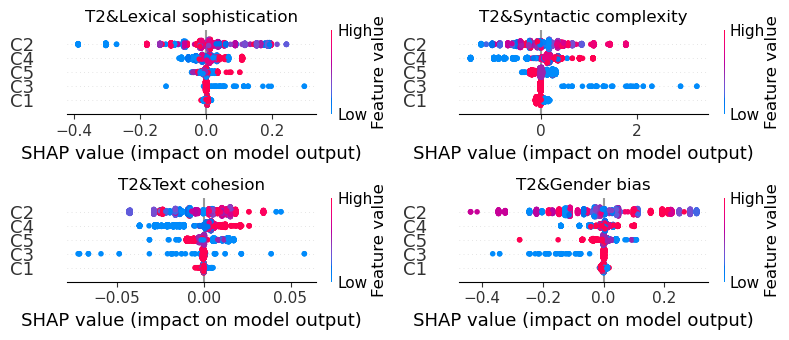}
\caption{Beeswarm plot of \texttt{seek suggestions -> accept without revision} (T2) \& outcomes.} 
\label{fig3}
\end{figure*}

\begin{figure*}[hbt!]
\includegraphics[width=0.65\textwidth]{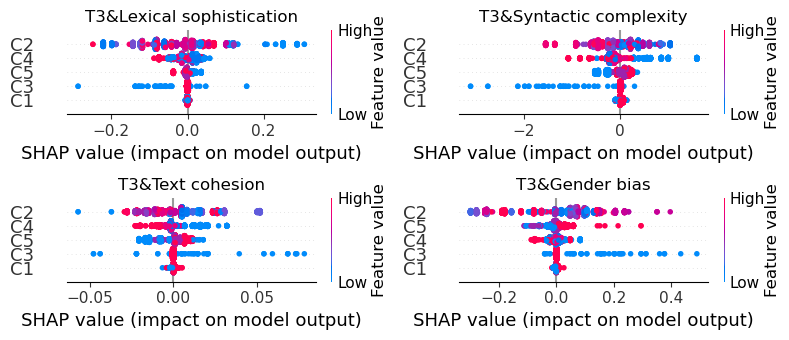}
\caption{Beeswarm plot of \texttt{seek suggestions -> first accept and the revise} (T3) \& outcomes.} 
\label{fig4}
\end{figure*}

\begin{table*}[hbt!]
\begin{center}
\caption{Summary of the impact of the writing behavior on outcomes by confounders. A '-' indicates no clear trend observed, '$\uparrow$' signifies a positive impact, and '$\downarrow$' denotes a negative impact. The arrow with a star indicates that the trend in the sub-groups contradict the overall ATE trend shown in Table \ref{table2}.}
\label{table3}
\resizebox{0.65\textwidth}{!}{
\begin{tabular}{@{}cccccccccccccc@{}}
\toprule
\multirow{2}{*}{Confounder} & \multirow{2}{*}{Confounder Value} & \multicolumn{3}{c}{\begin{tabular}[c]{@{}c@{}}Y1: Lexical \\ sophistication\end{tabular}} & \multicolumn{3}{c}{\begin{tabular}[c]{@{}c@{}}Y2: Syntactic \\ complexity\end{tabular}} & \multicolumn{3}{c}{\begin{tabular}[c]{@{}c@{}}Y3: Text \\ cohesion\end{tabular}} & \multicolumn{3}{c}{\begin{tabular}[c]{@{}c@{}}Y4: Gender \\ bias\end{tabular}} \\ \cmidrule(l){3-14} 
 &  & T1 & T2 & \multicolumn{1}{c|}{T3} & T1 & T2 & \multicolumn{1}{c|}{T3} & T1 & T2 & \multicolumn{1}{c|}{T3} & T1 & T2 & T3 \\ \midrule
\multirow{2}{*}{C1: Writing genre} & Argumentative & $\stackrel{*}{\uparrow}$ & - & \multicolumn{1}{c|}{-} & - & - & \multicolumn{1}{c|}{-} & - & $\stackrel{*}{\uparrow}$ & \multicolumn{1}{c|}{$\stackrel{*}{\downarrow}$} & - & - & - \\ \cmidrule(lr){2-2}
 & Creative & $\downarrow$ & - & \multicolumn{1}{c|}{-} & - & - & \multicolumn{1}{c|}{-} & - & $\downarrow$ & \multicolumn{1}{c|}{$\uparrow$} & - & - & - \\ \cmidrule(r){1-2}
\multirow{2}{*}{C3: Language background} & Non-native & $\stackrel{*}{\uparrow}$ & $\stackrel{*}{\uparrow}$ & \multicolumn{1}{c|}{-} & - & $\stackrel{*}{\uparrow}$ & \multicolumn{1}{c|}{-} & $\stackrel{*}{\downarrow}$ & - & \multicolumn{1}{c|}{-} & $\stackrel{*}{\uparrow}$ & $\downarrow$ & $\stackrel{*}{\uparrow}$ \\ \cmidrule(lr){2-2}
 & Native & - & - & \multicolumn{1}{c|}{-} & - & - & \multicolumn{1}{c|}{-} & - & - & \multicolumn{1}{c|}{-} & - & - & - \\ \cmidrule(r){1-2}
\multirow{2}{*}{C4: GPT temperature} & Low & - & - & \multicolumn{1}{c|}{-} & $\downarrow$ & - & \multicolumn{1}{c|}{-} & - & $\downarrow$ & \multicolumn{1}{c|}{$\uparrow$} & - & - & - \\ \cmidrule(lr){2-2}
 & High & - & - & \multicolumn{1}{c|}{-} & $\stackrel{*}{\uparrow}$ & - & \multicolumn{1}{c|}{-} & - & $\stackrel{*}{\uparrow}$ & \multicolumn{1}{c|}{$\stackrel{*}{\downarrow}$} & - & - & - \\ \cmidrule(r){1-2}
\multirow{2}{*}{C5: GPT frequency penalty} & Low & - & - & \multicolumn{1}{c|}{-} & $\downarrow$ & - & \multicolumn{1}{c|}{$\stackrel{*}{\downarrow}$} & - & - & \multicolumn{1}{c|}{-} & - & $\stackrel{*}{\uparrow}$ & - \\ \cmidrule(lr){2-2}
 & High & - & - & \multicolumn{1}{c|}{-} & $\stackrel{*}{\uparrow}$ & - & \multicolumn{1}{c|}{-} & - & - & \multicolumn{1}{c|}{-} & $\downarrow$ & $\downarrow$ & - \\ \bottomrule
\end{tabular}
}
\end{center}
\end{table*}

\subsection{Results on Average Treatment Effect}
The results of ATE estimations and the refutation tests for each pair of treatment and outcome are presented in Table \ref{table2}, which provides insight into the overall causal relationship between treatments and outcomes across the entire population. As shown in Table \ref{table2}, all the causal models for each treatment-outcome pair successfully passed the three refutation tests (i.e., RCC, Placebo, and DSR), with all p-values exceeding 0.05. This indicates that our causal inference results are both robust and credible.

T1 (seek suggestions -> not accept) was associated with a reduction in both Y1 (lexical sophistication) by 0.065 and Y2 (syntactic complexity) by 0.236, but it led to a slight increase in Y3 (text cohesion) by 0.001. In comparison, T2 (seek suggestions -> accept without revision) exhibited the most substantial negative impact across all three quality measures, reducing Y1 by 0.086, Y2 by 0.869, and Y3 by 0.007. Interestingly, T3 (seek suggestions -> first accept and the revise) improved all three quality metrics, with increases in Y1 by 0.102, Y2 by 0.963, and Y3 by 0.008. These results suggest that writers who frequently rely on GAI to write for them exhibit the poorest performance in terms of writing quality. In contrast, writers who use GAI primarily for ideation but compose their own text perform better, particularly showing a positive impact on text cohesion. Writers who actively revise GAI suggestions achieve the highest performance in writing quality, indicating that the process of critically engaging with and refining GAI-generated suggestions enhances lexical sophistication, syntactic complexity, and text cohesion in the written work. For Y4 (gender bias), all three GAI-assisted writing behaviors result in a decrease in gender bias. T1 shows the largest decrease (about 0.163), compared to smaller reductions for T2 (0.012) and T3 (0.013). This appears reasonable, as T1 does not directly incorporate GAI-generated text, while T2 and T3 both involve accepting GAI-generated content, whether modified or not. This aligns with existing research suggesting that LLMs often contain biases. Writers who modify GAI suggestions can critically assess and adjust biased language, making it slightly better than directly accepting GAI suggestions.

\subsection{Results on Individual Treatment Effect}
To explore the effect of treatments on outcomes at the individual level (i.e., ITE), we visualized the Beeswarm plots for each treatment, conditioned on different confounders. Figure \ref{fig2} shows this for T1, Figure \ref{fig3} for T2, and Figure \ref{fig4} for T3. To further highlight the impact of treatments on outcomes by confounders, summary results are provided in Tables \ref{table3}. Only clear patterns (i.e., where most dots of the same color accumulate on either the left or right of the y-axis) observed in the Beeswarm plots were included in the summary table. These patterns were empirically selected and verified by three authors. Note that C2, containing 20 different writing topics, did not exhibit clear trends and is therefore omitted.

When analyzing T1's ITE on outcomes by confounders, as presented in Figure \ref{fig2}, we are essentially examining how seeking external GAI suggestions during the planning stage, without accepting those suggestions, influences the quality of the final product across various confounder groups. For T2's ITE on outcomes by confounders, shown in Figure \ref{fig3}, we are investigating how directly incorporating AI-generated sentences during the translation stage affects the final product's quality across different confounder groups. Lastly, in the case of T3's ITE on outcomes by confounders, displayed in Figure \ref{fig4}, we are exploring how writers refining AI-generated text to better align with their tone or content during the reviewing stage impacts the quality of the final product across various confounder groups. As demonstrated in Table \ref{table3}, we observed several noteworthy findings, especially within certain confounder groups, where we even identified trends that contradicted (i.e., indicated by arrows with stars in the table) the ATE (i.e., the overall average causal effect across the entire population).

For Y1 (lexical sophistication), we first observed that seeking GAI suggestions without directly accepting them (T1) has a positive effect on lexical sophistication in argumentative writing, while the effect is negative in creative writing. A possible reason is that GAI often produces more logical and formal text, which aligns with the demands of argumentative writing. Writers may be inspired by GAI suggestions to use more advanced vocabulary, even if they do not directly accept the suggestions. Additionally, we found that non-native English writers tended to benefit from either accessing (T1) or directly adopting GAI-generated text (T2). This may be because non-native writers often have a smaller vocabulary compared to their native English counterparts, and exposure to or direct use of advanced vocabulary in GAI suggestions may enhance the lexical sophistication of their written products.

For Y2 (syntactic complexity), we first observed that directly adopting GAI-generated text without modification (T2) positively impacts syntactic complexity for non-native English writers. This aligns with our findings on Y1, where non-native writers may tend to use simpler syntactic structures compared to their native counterparts. By incorporating GAI-generated text into their essays, non-native writers can enhance the syntactic complexity of their final products. We also found that in high GPT temperature and frequency penalty settings, writers who frequently seek GAI suggestions without directly accepting them (T1) tend to produce more syntactically complex essays. One possible explanation is that GPT-generated text introduces varied vocabulary and greater creativity in these settings, inspiring writers to produce more complex structures in their own writing. Lastly, we found that writers who prefer to accept and revise GAI suggestions (T3) may produce essays with lower syntactic complexity in low GPT frequency penalty settings. This highlights that even writers who demonstrate meaningful learning by actively revising GAI-generated content may still produce essays with less syntactic complexity in these settings. Therefore, assessing only the final written products in GAI-assisted writing tasks is insufficient. Educators need to incorporate new assessment methods to better support students in such contexts.

For Y3 (text cohesion), we first observed that writers who frequently accept GAI suggestions without further modification (T2) tend to produce more cohesive text compared to those who actively modify GAI suggestions (T3) in argumentative writing tasks. One possible reason is that GAI can generate logical and formal text that aligns with the requirements of argumentative writing. Consequently, T2 benefits from incorporating such text directly into their essays, while T3 may reduce the effectiveness of GAI suggestions through modifications. This underscores our argument that meaningful learning during the writing process may not always result in cohesive essays for certain confounding groups, indicating that assessing only the final written product in GAI-assisted tasks is insufficient. We also found that non-native writers may reduce text cohesion if they frequently seek GAI suggestions without directly accepting them (T1). A potential explanation is that although these writers, who may face more difficulty writing in English, gain inspiration from GAI-generated content, they still struggle to produce cohesive text on their own. Lastly, when GPT temperature is high, meaning the GAI-generated text becomes more diverse and creative, text cohesion tends to benefit less from active revisions of GAI suggestions. This may be due to the difficulty of revising creative, varied text to improve overall cohesion in an essay.

For Y4 (gender bias), we first found that non-native English writers who directly use GAI-generated text (T2) tend to produce essays with lower gender bias compared to those written primarily on their own (T1) or those consisting of revised GAI-generated text (T3). This suggests that non-native writers may unintentionally introduce biased language, highlighting the need for additional support to help them avoid biased terms in their writing. We also found that in low GPT frequency penalty settings, the direct use of GAI-generated text (T2) may result in less biased essays. This could be because low frequency penalty settings often lead to less varied vocabulary, potentially reducing the likelihood of generating biased terms.

\section{Discussion and Conclusion}
In this study, we used causal modeling to identify causal relationships within an observational GAI-assisted writing dataset. We defined three distinct treatments: T1 (seek suggestions -> not accept), T2 (seek suggestions -> accept without revision), and T3 (seek suggestions -> first accept and the revise). Additionally, we identified four outcome measures related to the written outputs: Y1 (lexical sophistication), Y2 (syntactic complexity), Y3 (text cohesion), and Y4 (gender bias). To control for potential confounding variables, we considered five confounders: C1 (writing genre), C2 (writing topic), C3 (language background), C4 (GPT temperature), and C5 (GPT frequency penalty). We applied the state-of-the-art X-learner algorithm to infer causal relationships and analyzed both the ATE and ITE.

\noindent \textbf{Discussion.} The ATE results show that T3 consistently and significantly improves all writing quality measures (Y1, Y2, and Y3), while T2 tends to significantly reduce all of these measures. T1 slightly improves Y3, but reduces both Y1 and Y2. The positive effect of T3 suggests that writers who actively engage with AI-generated content by critically refining GAI suggestions produce writing with more sophisticated vocabulary, complex sentence structures, and cohesive content. These findings are consistent with prior research, which suggests that deeper interaction with AI tools fosters critical thinking and creativity in language use \cite{putjorn2023augmented}. These results also support the argument that the review phase is critical for improving text quality and fostering learning, as highlighted in traditional writing research \cite{van1997review, baker2016peer}. Interestingly, the negative effects of T1 and T2 suggest that over-reliance on GAI suggestions for generating ideas and text may be detrimental to students' learning. This could be because GAI reduces the need for writers to brainstorm ideas and construct complex sentences on their own. This reinforces the notion that GAI may hinder active linguistic engagement when writers depend too heavily on GAI \cite{abdelghani2023generative}. Based on these findings, we argue that students' GAI-assisted writing behaviors can influence the quality of written products, with the final product reflecting not only their writing abilities but also how they use GAI. Therefore, educators should focus not only on the written product in GAI-assisted writing but also on students' engagement with GAI (e.g., whether they simply accept GAI suggestions or thoughtfully revise them) during the writing process and how this engagement impacts the writing quality. Furthermore, educators should train students to critically refine AI-generated suggestions throughout the writing process, helping them adapt these suggestions to align with their own voice and ideas.

\noindent \textbf{Implications.} The ITE results provide several important implications. Firstly, we argue that linguistic bias in written text has not been thoroughly explored in previous literature, particularly in the context of GAI-assisted writing tasks, and warrants further investigation. Our findings indicate that all three GAI-assisted writing behaviors contribute to a reduction in gender bias, suggesting that linguistic bias is already present in students' writing, even without GAI assistance. Notably, non-native English writers are more likely to unintentionally amplify gender biases when modifying GAI suggestions or writing independently, likely due to unfamiliarity with nuanced gendered language. Additionally, the reduction in bias is significantly greater in T1 compared to T2 and T3, which is expected, as T1 does not directly incorporate GAI-generated text, whereas T2 and T3 do. This underscores the importance of designing targeted training for educators to address bias in both GAI outputs and students' writing, especially for non-native writers. Secondly, students with different demographic attributes require tailored support in various writing and GAI settings to achieve meaningful learning through GAI-assisted writing practices. For instance, writers who demonstrated meaningful learning by actively modifying GAI suggestions in low GPT temperature settings often produced essays with lower syntactic complexity. Educators should recognize this and offer targeted guidance to help these writers effectively adjust GAI suggestions, ensuring that their revisions align with their own voice and ideas while enhancing the syntactic complexity of their writing. Similarly, in argumentative writing, students who displayed meaningful learning often produced essays with lower text cohesion. Educators should also provide guidance on how to effectively modify GAI suggestions in argumentative writing tasks to help students create more cohesive essays. Thirdly, there is a need to develop new in-process assessment methods for GAI-assisted writing to better evaluate students' writing performance and enhance meaningful learning. It is evident that some non-native writers, who showed less meaningful learning by frequently accepting GAI suggestions without modification, still produced high-quality essays. Similarly, in high GPT temperature settings, students who exhibited less meaningful learning also managed to create high-quality essays. This highlights that evaluating only the final written products in GAI-assisted writing tasks is insufficient and may cause some students to miss valuable learning opportunities during the writing process.

\noindent \textbf{Limitations.} We acknowledge several limitations in our study. Firstly, our findings are based on the CoAuthor dataset, which may limit their generalizability to other writing contexts with different subjects or requirements. In future work, we aim to gather additional data from more diverse writing settings to further validate our results. Secondly, the limited availability of writer-related information, such as GAI literacy, restricts the inclusion of relevant confounders. To address this, we plan to incorporate surveys or tests in future data collection efforts to better capture these factors. Thirdly, the imbalance in data distribution (e.g., no-native writers vs. native writers) within the dataset may have influenced the results. To mitigate this, we intend to recruit a more balanced group of writers in future studies.

\begin{acks}
This research was funded partially by the Australian Government through the Australian Research Council (project number DP240100069 and DP220101209) and the Jacobs Foundation (CELLA 2 CERES). 
\end{acks}

\bibliographystyle{ACM-Reference-Format}
\bibliography{sample-base}

\end{document}